\documentclass[12pt,epsf]{article}
\usepackage{amsmath}
\usepackage{amssymb}
\usepackage{bm}
\usepackage[dvips]{graphicx}
\begin{document}
\setlength{\oddsidemargin}{0cm}
\setlength{\baselineskip}{7mm}

\begin{titlepage}

\begin{center}
  {\LARGE
Multi-cut Solutions in Chern-Simons Matrix Models
  }
\end{center}
\vspace{0.2cm}
\baselineskip 18pt 
\renewcommand{\thefootnote}{\fnsymbol{footnote}}

\begin{center}
Takeshi {\sc Morita}\footnote{%
E-mail address: morita.takeshi@shizuoka.ac.jp
} and 
Kento S{\sc ugiyama}\footnote{
E-mail address: sugiyama.kento.15@shizuoka.ac.jp}

\renewcommand{\thefootnote}{\arabic{footnote}}
\setcounter{footnote}{0}

\vspace{0.4cm}

{\small\it

Department of Physics,
Shizuoka University, \\
836 Ohya, Suruga-ku, Shizuoka 422-8529, Japan

}

\end{center}


\vspace{1.5cm}

\begin{abstract}

We elaborate the Chern-Simons (CS) matrix models at large $N$.
The saddle point equations of these matrix models have a curious structure which cannot be seen in the ordinary one matrix models.
Thanks to this structure, an infinite number of multi-cut solutions exist in the CS matrix models.
Particularly we exactly derive the two-cut solutions at finite 't\,Hooft coupling in the pure CS matrix model.
In the ABJM matrix model, we argue that some of multi-cut solutions might be interpreted as a condensation of the D2-brane instantons.

\end{abstract}


\end{titlepage}

\section{Introduction}

Matrix models play crucial roles in various topics in theoretical physics. 
(See, for example, recent review \cite{Eynard:2015aea, Marino:2012zq}.)
Particularly, in string theory, there are several proposals that matrixes may provide the non-perturbative formulation of non-critical strings \cite{Brezin:1990rb, Douglas:1989ve, Gross:1989vs, Gross:1989aw} and critical strings \cite{Banks:1996vh, Ishibashi:1996xs,Dijkgraaf:1997vv}, and, many remarkable results have been obtained in this direction.
Besides, since matrixes are also related to the large-$N$ gauge theories, they play special roles in quantum gravity through the gauge/gravity correspondence \cite{Maldacena:1997re, Itzhaki:1998dd, Berenstein:2002jq, Aharony:2008ug}.

Recently, the importance of matrix models is significantly increasing in supersymmetric gauge theories and related mathematical physics too.
In supersymmetric theories, the dynamical degree of freedom drastically reduce due to the cancellation between the bosons and fermions, and, in some special cases, the theories are effectively described by zero dimensional matrix models \cite{Nekrasov:2002qd, Dijkgraaf:2002dh, Cachazo:2002ry, Nekrasov:2003rj}.
Especially, through the development of the localization technique \cite{Pestun:2007rz, Kapustin:2010xq}, many correspondences between the supersymmetric gauge theories and matrix models have been found. (See reviews \cite{Marino:2011nm, Hosomichi:2015jta, Pestun:2016zxk}.)

Among these matrix models, we focus on the $U(N)$ Chern-Simons (CS) matrix models in this article.
The partition function of the pure CS matrix model is given by \cite{Marino:2002fk, Aganagic:2002wv, Kapustin:2009kz}  
\begin{equation}
Z_N^{\rm CS}(k)
=\frac{1}{N!} \int \prod_i \frac{du_i}{2\pi} e^{-\frac{N}{4\pi i \lambda} \sum_i u^2_i} \prod_{i<j} \Bigl[ 2 \sinh{\frac{u_i-u_j}{2}} \Bigr]^2.
\label{partition}
\end{equation}
Here $\lambda \equiv N/k$ is the 't\,Hooft coupling and $k$ is an integer representing the CS level.
The computation of this integral has been already done  and we know the exact value of this partition function as a function of $k$ and $N$ \cite{Kapustin:2009kz, Tierz:2002jj}.
However investigating this model is still valuable.
As we will see, this model shows curious properties which have not been seen in any other one matrix models.
In addition, the varieties of the pure CS matrix model are being actively studied, and understanding this model may help us in developing insights into these models.
For example, the ABJM theory \cite{Aharony:2008ug} on $S^3$ is described by a similar matrix integral \cite{Kapustin:2009kz}.
Also the three dimensional ${\mathcal N}=2$ supersymmetric CS matter theory coupled to matters with arbitrary $R$-charge on $S^3$ is described by a related matrix model \cite{Jafferis:2010un, Hama:2010av}.
Especially the 't\,Hooft limit of these models are important in string theory, and we study these CS matrix models under this limit.

\begin{figure}
\begin{tabular}{cc}
\begin{minipage}{0.5\hsize}
\begin{center}
        \includegraphics[scale=0.6]{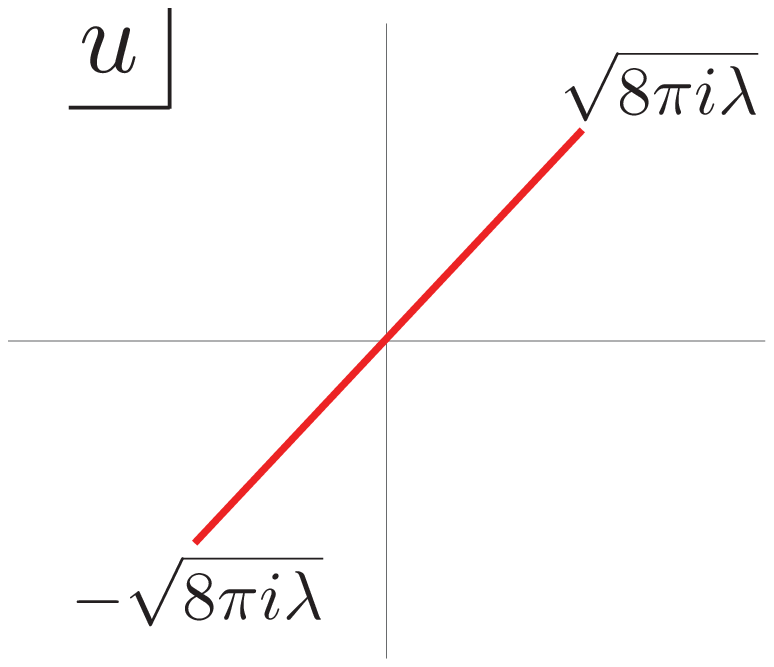}\\
    one-cut solution
\end{center}
\end{minipage}
\begin{minipage}{0.5\hsize}
\begin{center}
        \includegraphics[scale=0.55]{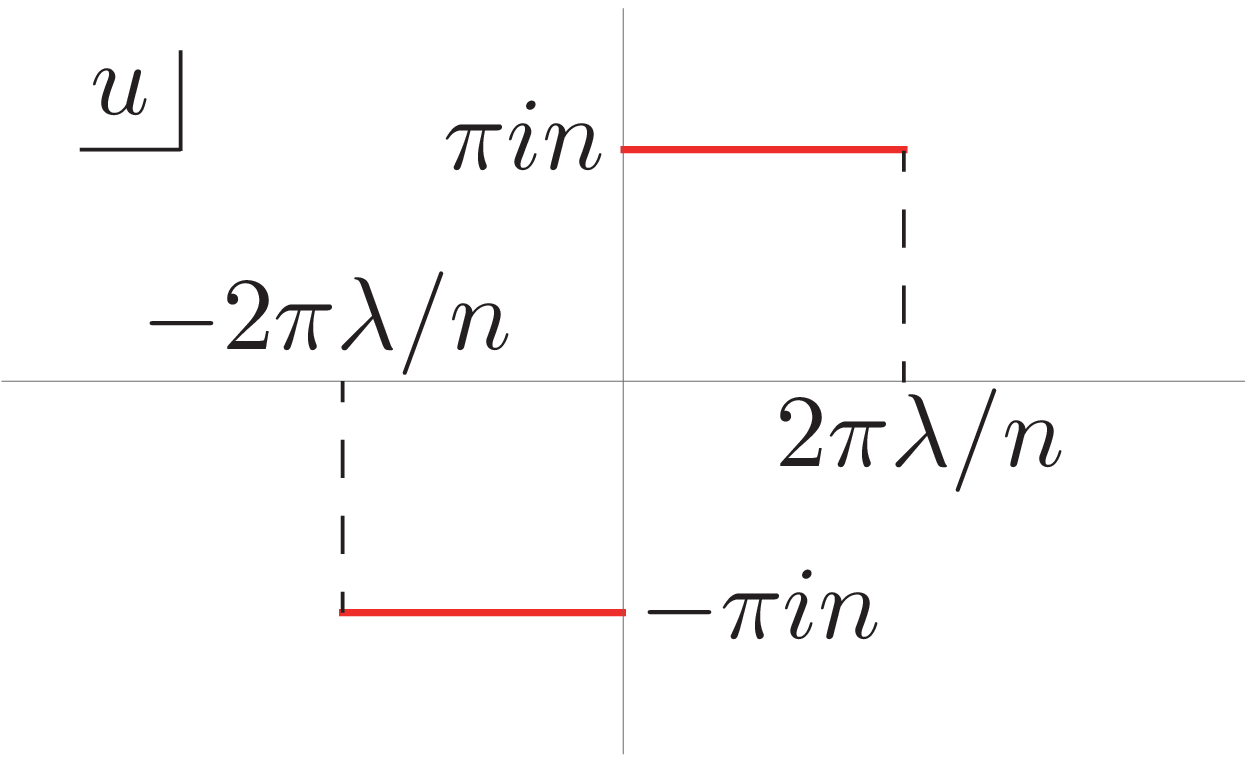}\\
        two-cut solution
\end{center}
\end{minipage}
\end{tabular}
       \caption{ Solutions of the pure CS matrix model at weak coupling $|\lambda| \ll 1$.
       The red lines represent the eigenvalue distributions (cuts).
       $n$ in the two-cut solution is an integer.
       }
        \label{fig-sol-CS}
\end{figure}

When we take the 't\,Hooft limit ($N \to \infty$, $\lambda$ fixed), the saddle point approximation may be applicable.
The saddle point equation of the pure CS matrix model (\ref{partition}) is given by
\begin{equation}
u_i=\frac{2\pi i \lambda}{N} \sum_{j \neq i} \coth{\frac{u_i-u_j}{2}}, \qquad (i=1,\cdots, N).
\label{eom-u}
\end{equation}
The solution of this equation at finite $\lambda$  has been found, and it is characterized by a single cut of the eigenvalue distribution \cite{Aganagic:2002wv, Halmagyi:2003ze, Pasquetti:2009jg}.
(See figure \ref{fig-sol-CS} (left).)

Then a natural question is whether the solution is unique.
We find that the answer is no.
The saddle point equation (\ref{eom-u}) has an interesting structure which allows an infinite number of solutions characterized by multi-cut eigenvalue distributions. 
See figure \ref{fig-sol-CS} (right) for a two-cut solution.
Moreover these multi-cut solutions would ubiquitously exist in the varieties of the CS matrix models too.
(Actually such multi-cut solutions were first found in the CS matrix model coupled to adjoint matters through a numerical analysis \cite{Morita:2011cs}.)

In one matrix models, the multi-cut solutions are usually related to the large-$N$ instantons \cite{David:1990sk, David:1992za}.
We argue that indeed some of the multi-cut solutions in the ABJM theory might be interpreted as a condensation of the D2-brane instantons \cite{Drukker:2011zy}.
Besides, some of the multi-cut solutions in the CS matrix model coupled to adjoint matter case might be related to the poles in the Borel plane \cite{Honda:2016vmv}.
However, we have not understood the physical interpretations of other multi-cut solutions.
They might be related to some non-perturbative objects in the CS matrix models and string theory. 

Although we find the new multi-cut solutions in the CS matrix models, 
 we have to emphasize that we just evaluate the saddle point equation (\ref{eom-u}).
It means that we cannot answer whether these solutions should be summed up through the path integral (\ref{partition}).
This is a crucial question but we leave this problem for future work.\\

The organization of this paper is as follows:
In section \ref{sec-CS}, we explain the basic idea of the derivation of the multi-cut solutions in the pure CS matrix model at weak coupling.
(We show the derivation at finite coupling in appendix \ref{app-Exact}.)
In section \ref{sec-Ad}, we consider the CS matrix model coupled to adjoint matters.
There, we argue the relation to the poles in the Borel plane in this model.
In section \ref{sec-ABJM}, we argue the multi-cut solutions in the ABJM matrix model and show the connection to the D2-brane instantons.
Section \ref{sec-discussion} list some discussions for future explorations.


\section{Multi-cut Solutions in pure CS matrix model}
\label{sec-CS}

\subsection{One-cut solution}

We first briefly review how the one-cut solution  in the pure CS matrix model (\ref{partition}) which has been investigated in the previous studies is obtained.
To see it quickly, we take a weak coupling limit $ |\lambda| \ll 1$ of the saddle point equation (\ref{eom-u}), although we can solve it even at finite $\lambda$  \cite{Aganagic:2002wv, Halmagyi:2003ze, Pasquetti:2009jg}.
Under this limit, the quadratic potential of the path integral (\ref{partition}) dominates and the eigenvalues may be strongly confined around $u_i \simeq 0$.
Thus we can approximate the right hand side of the saddle point equation (\ref{eom-u}) as
\begin{equation}
V'_{\rm G} (u_i)=\frac{2\lambda}{N} \sum_{j \neq i} \frac{1}{u_i-u_j},  
\qquad (i=1,\cdots N), \qquad V_{\rm G} (u) \equiv \frac{1}{4 \pi i } u^2.
\label{eom-hermite}
\end{equation}
This is the well-known saddle point equation for the Hermitian matrix model with a (complex) Gaussian potential, and we can solve this equation at large $N$  by using the resolvent.
Although the derivation of the solution is well known, since we will use this technique to obtain the multi-cut solutions too, we briefly explain it here.

To solve equation (\ref{eom-hermite}) at large $N$, we introduce the eigenvalue density $\rho(z)$ and the resolvent $v(z)$ which are defined by
\begin{align}
\rho(z) = \frac{1}{N} \sum_i^N \delta(z-u_i) , \qquad v(z)=\int_a^b dw  \frac{\rho(w)}{z-w}.
\label{resolvent-Her}
\end{align}
Here $a$ and $b$ denote the location of the end points of the eigenvalue distribution, which will be determined later.
By using the resolvent, the saddle point equation (\ref{eom-u}) becomes\footnote{ $\epsilon $ may be a complex, since the cut does not need to be on the real axis in the CS matrix models.}
\begin{align}
V_{\rm G}'(z) =  \lambda \lim_{\epsilon \rightarrow 0} \left[ v(z+i\epsilon)+v(z-i\epsilon) \right] ,\qquad (z \in [a,b] ~) 
\label{eq-resolvent}
\end{align}
and the eigenvalue density can be read off from the resolvent 
\begin{align}
\rho(z)
=& -\frac{1}{2\pi i} \lim_{\epsilon \rightarrow 0} \left[ v(z+i\epsilon)-v(z-i\epsilon) \right] ,\qquad (z \in [a,b] ~) .
\end{align}
Also, through the definition of the resolvent (\ref{resolvent-Her}), $v(z)$ shows the asymptotic behavior
\begin{align}
v(z) \rightarrow \frac{1}{z}, \qquad (z \to \infty).
\label{boundary-v}
\end{align}
We will use this equation as the boundary condition of $v(z)$.

The one-cut solution of equation (\ref{eq-resolvent}) can be obtained by using the following ansatz \cite{Migdal:1984gj}
\begin{align}
v(z)=\frac{1}{\lambda} \int_{C_1} \frac{dw}{4\pi i} \frac{V_{\rm G}'(w)}{z-w} \sqrt{\frac{(z-a)(z-b)}{(w-a)(w-b)}},
\end{align}
where the integral contour $C_1$  goes around the cut$:[a,b]$ in a counter-clockwise way.
By performing this integral and imposing the boundary condition (\ref{boundary-v}), we obtain
\begin{align}
v(z) = \frac{z}{ 4 \pi i \lambda} - \frac{1}{4 \pi i \lambda} \sqrt{z^2 - 8 \pi i \lambda},
\label{one-cut-sol}
\end{align}
where $a$ and $b$ have been determined $b=-a= \sqrt{8 \pi i \lambda} $.
Obviously this resolvent satisfies the equation (\ref{eq-resolvent}).
From this result, we obtain the eigenvalue density,
\begin{align}
\rho(z) = \frac{1}{4 \pi^2 i \lambda} \sqrt{ 8 \pi i \lambda  - z^2}, \qquad \left(z \in \left[-\sqrt{8 \pi  i \lambda},\sqrt{8 \pi  i \lambda}\right] \right).
\label{one-cut-rho}
\end{align}
Thus the eigenvalues are distributed in the complex plane as shown in figure \ref{fig-sol-CS} (left) \footnote{In addition to this cut, its images appears on $z \in \left[-\sqrt{8 \pi  i \lambda} +2 \pi in,\sqrt{8 \pi  i \lambda} +2 \pi in \right]$, ($n \in \mathbf Z $) due to the periodicity of the coth in the saddle point equation (\ref{eom-u}). 
In this paper, we do not count these images as the number of the cuts.}.

\subsection{Two-cut solution}
Now we show that the saddle point equation (\ref{eom-u}) even at weak coupling has a non-trivial structure and other solutions exist.
As a warm up, we first consider the saddle point equation (\ref{eom-u}) at $N=2$.
\begin{equation}
\frac{u_1}{2\pi i \lambda}=\frac{1}{2} \coth{\frac{u_1-u_2}{2}},  \qquad
\frac{u_2}{2\pi i \lambda}=-\frac{1}{2} \coth{\frac{u_1-u_2}{2}}.
\end{equation}
By summing these equations, we immediately find $u_2=-u_1$ and the equations reduce to
\begin{equation}
\frac{u_1}{2\pi i \lambda}=\frac{1}{2} \coth u_1.
\end{equation}
Obviously this equation has an infinite number of solutions.
If $\lambda$ is small, we can perturbatively solve this equation and obtain
\begin{align}
u_1&= \pm \sqrt{\pi i \lambda} + \cdots , \\
u_1&= \pi i n + \frac{\lambda}{n}  + \cdots , 
\label{N=2-multi}
\end{align}
where $n$ is a non-zero integer.
If we take $N$ large, the first equation would correspond to the one-cut solution (\ref{one-cut-rho}).
On the other hand, the second solution indicates the existence of another class of the solutions\footnote{A related speculation was first done in Ref.\cite{Morita:2011cs}.}.

Let us try to find the solution corresponding to the second solution at large $N$. By regarding the result (\ref{N=2-multi}), it is natural to take the following ansatz,
\begin{align}
u_i
&=   \pi i n   +x_i ,  \qquad (i=1, \cdots , N/2),
\nonumber \\
u_{N/2+j}&
= -  \pi i n +y_j ,\quad (j=1, \cdots , N/2).
\label{ansatz-2-cut}
\end{align}
We can choose the integer $n$ positive without loss of generality.
By substituting this ansatz into the saddle point equation (\ref{eom-u}), we obtain
\begin{align}
 \frac{n}{2} +\frac{1}{2\pi i} x_i & =\frac{\lambda}{N} \sum_{ j \neq i}^{N/2} \coth{\frac{x_i-x_j}{2}} + \frac{\lambda}{N} \sum_{j =1}^{N/2} \coth{\frac{x_i-y_j}{2}}, \qquad (i=1, \cdots , N/2) ,\nonumber \\
-  \frac{n}{2}+\frac{1}{2\pi i} y_i & =\frac{\lambda}{N} \sum_{ j \neq i}^{N/2} \coth{\frac{y_i-y_j}{2}} + \frac{\lambda}{N} \sum_{j =1}^{N/2} \coth{\frac{y_i-x_j}{2}} ,\qquad (i=1, \cdots , N/2).
\label{eom-finite}
\end{align}
From now, we focus on the weak coupling limit $|\lambda | \ll 1$ and consider the leading order in the expansion.
(We show the result at finite $\lambda$ in appendix \ref{app-Exact}.)
We assume that the eigenvalues locate around $x_i = 0$ and $y_j = 0$ at weak coupling similar to the one-cut solution in the previous section. (Actually we will later see that $x_i, ~y_j \sim O(\lambda)$ similar to the $N=2$ case (\ref{N=2-multi}).)
Then equation (\ref{eom-finite}) reduce to
\begin{align}
\frac{n}{2} & =\frac{2\lambda}{N} \sum_{ j \neq i}^{N/2} \frac1{x_i-x_j} + \frac{2\lambda}{N} \sum_{j =1}^{N/2} \frac1{x_i-y_j},  \qquad (i=1, \cdots , N/2), \nonumber \\
-\frac{n}{2}& =\frac{2\lambda}{N} \sum_{ j \neq i}^{N/2} \frac1{y_i-y_j} + \frac{2\lambda}{N} \sum_{j =1}^{N/2} \frac1{y_i-x_j}, \qquad (i=1, \cdots , N/2).
\label{eom-weak}
\end{align}

Here we argue how ``forces" act on the eigenvalues $\{ x_i \}$ and $\{ y_j \}$.
The expression of the right hand side of these equations are similar to the Vandermonde repulsive force between the eigenvalues in the Hermitian matrix model (\ref{eom-hermite}).
The difference is that they also work between $x_i$ and $y_j$ as if they locate nearby, although they are actually separated by $ 2 \pi i n $  in the complex plane.
The left hand side of equation (\ref{eom-weak}) is an external force acting on the eigenvalue $x_i$ and $y_j$.
It drags $x_i$ towards the left direction ($x_i \to - \infty$) and $y_j$ towards the right direction ($y_j \to  \infty$).
These directions of the forces can be read off from the potential of the CS matrix model (\ref{partition}) $u_i^2/4 \pi i $ too.
At $u_i =  \pi i n  + x_i $, this potential can be expanded as $u_i^2/4 \pi i= \pi i  n^2/4+  n  x_i/2  + \cdots$\, which is a linear potential for $x_i$ at the leading order. 
From the sign of this potential, we confirm that $x_i$ is indeed dragged towards $x_i \to - \infty$.

\begin{figure}
\begin{tabular}{cc}
\begin{minipage}{0.5\hsize}
\begin{center}
        \includegraphics[scale=0.6]{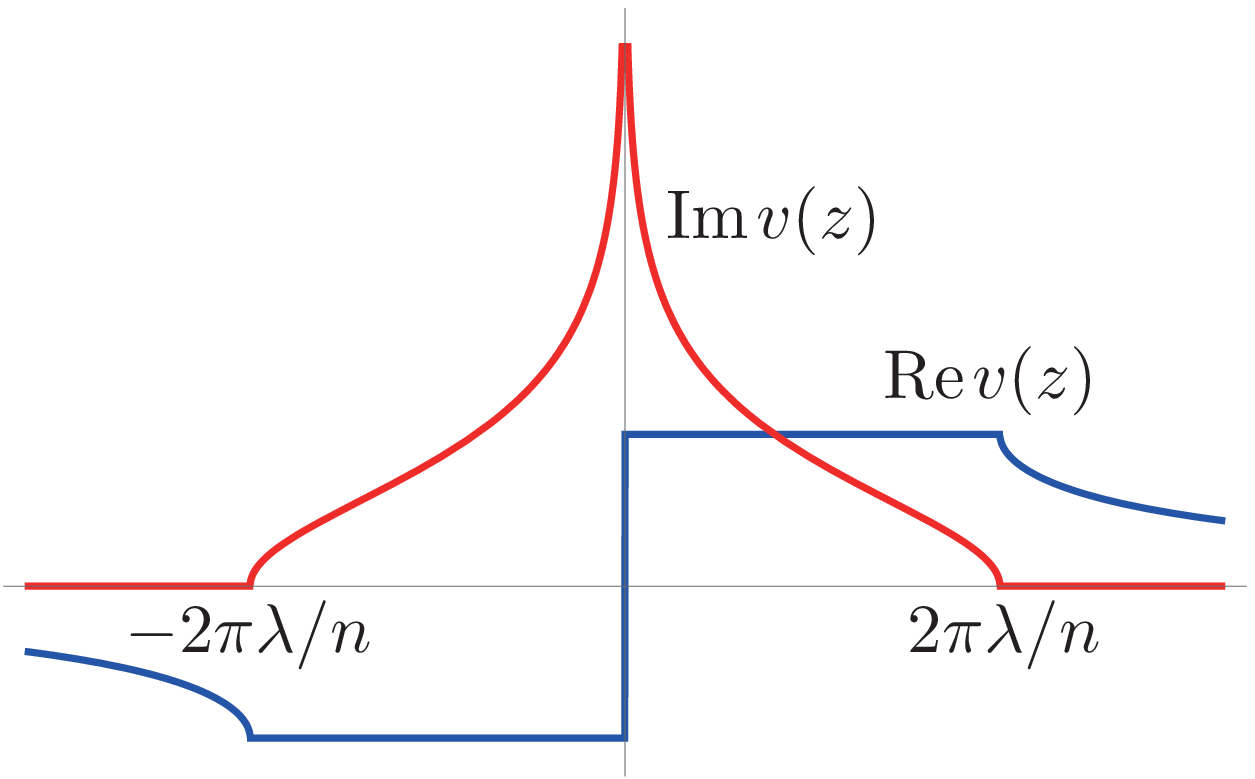}
\end{center}
\end{minipage}
\begin{minipage}{0.5\hsize}
\begin{center}
        \includegraphics[scale=0.5]{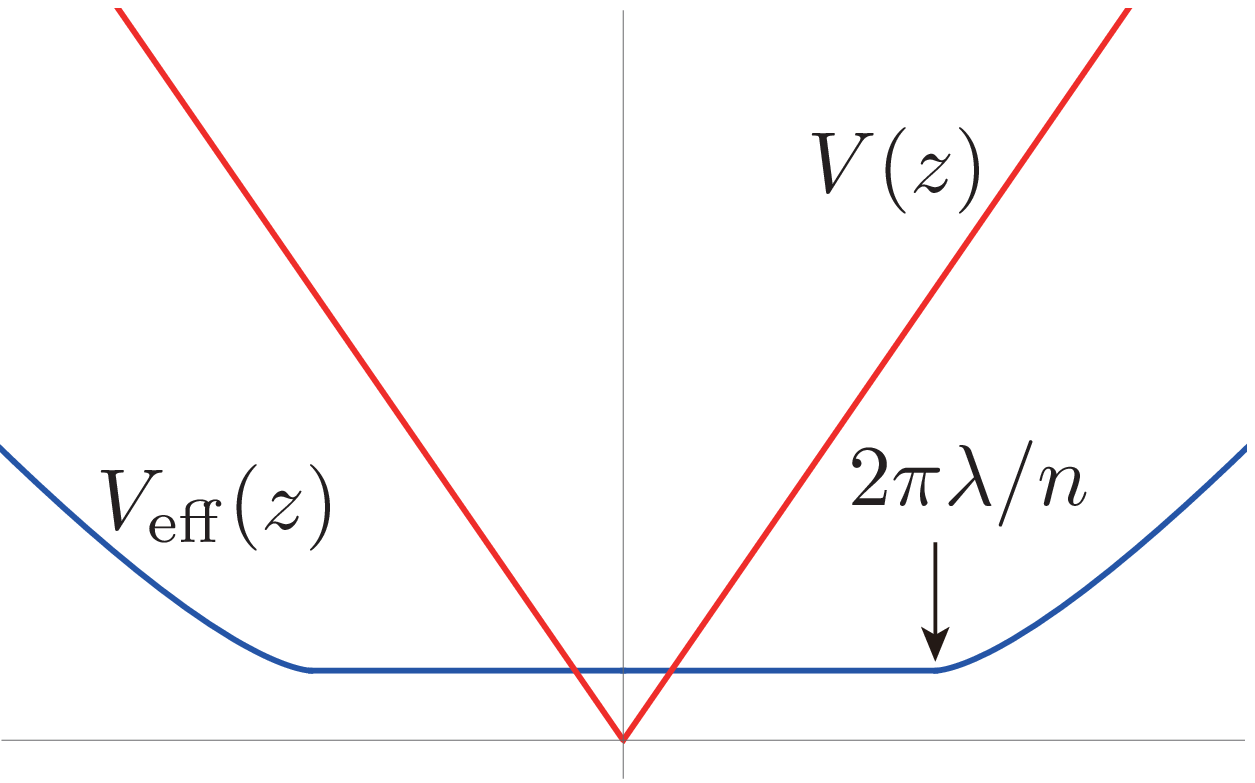}
\end{center}
\end{minipage}
\end{tabular}
        \caption{
(Left) Profile of the resolvent $v(z)$ (\ref{sol-two}). 
The red curve is Re$[v(z)]$ and  the blue curve is  Im$[v(z)]$  ($z \in {\mathbf R}$). 
 Re$[v(z)]$ on the cut agrees with the step function potential (\ref{eom-z-weak}) correctly.
        Im$[v(z)]/\pi $ describes the eigenvalue density $  \rho(z)$, and it shows an ``onion head" distribution. 
(Right)  The blue curve is the real part of the effective potential  $V_{\rm  eff}(z)$ for the single eigenvalue of the matrix in the two-cut solution. The red line is the classical potential $V(z)$ (\ref{potential-CS}).  
The plateau along the cut $[-2\pi \lambda/n, 2\pi \lambda/n]$ appears in the effective potential.
     }
        \label{fig-one-cut}
\end{figure}

Thus $x_i$ and $y_j$ tend to move toward left and right, respectively, and they repulse each other. 
Hence, if we set $x_i$ right and $y_j$ left, they may balance similar to the $N=2$ solution (\ref{N=2-multi}).
By regarding it, we assume that $x_i$ and $y_j$ satisfy
\begin{align}
{\rm Re}\left( y_j \right)  \le 0 \le {\rm Re}\left( x_i \right).
 \label{assumption-weak}
\end{align}
This assumption allows us to combine the saddle point equations for $x_i$ and $y_j$ (\ref{eom-weak}) into a single equation
\begin{align}
V'(z_i)  & =  \frac{2 \lambda}{N} \sum_{ j \neq i}^{N} \frac1{z_i-z_j}, \qquad
V'(z)=  n\left[ \theta \left({\rm Re}\left( z  \right)\right) - \frac{1}{2} \right], \qquad (i=1, \cdots , N),
\label{eom-z-weak}
\end{align}
where $z_i = x_i$ for $1 \le i \le N/2$ and $z_{i+N/2} = y_i$ for $1 \le i \le N/2$.
We have used the step function $\theta(x)$ in order to combine the left hand sides of the equation (\ref{eom-weak}).
Now the problem becomes much simpler.
This is merely a standard potential problem (\ref{eom-hermite}) with the singular potential 
\begin{align}
V(z)=   n z \left[  \theta \left({\rm Re}\left( z \right)\right) -  \frac{1}{2}\right].
\label{potential-CS}
\end{align}
We plot the profile of $V(z)$ in figure \ref{fig-one-cut} (right).
Although $V(z)$ is singular, this potential has the unique minimum at $z=0$ and the solution would be given by an ``one-cut" configuration. 
Here ``one-cut" means ``one-cut" of $z_i$ and it indeed describes the two-cuts of the original variable $u_i$ as shown in figure \ref{fig-sol-CS}.
Hence we introduce the resolvent $v(z)$ as in the previous section and apply the ansatz for the one-cut solution
\begin{align}
v(z)=&\frac{1}{\lambda} \int_{C_1} \frac{dw}{4\pi i} \frac{V'(w)}{z-w} \sqrt{\frac{(z-a)(z-b)}{(w-a)(w-b)}} \nonumber \\
=& 
\frac{1}{\lambda} \int_{0}^b \frac{dw}{2\pi i} \frac{n}{z-w} \sqrt{\frac{(z-a)(z-b)}{(w-a)(w-b)}}+
\frac{1}{\lambda} \int_{C_1} \frac{dw}{4\pi i} \frac{-\frac{n}{2} }{z-w} \sqrt{\frac{(z-a)(z-b)}{(w-a)(w-b)}} .
\label{ansatz-2-cut-resolvent}
\end{align}
Here we have defined $a$ and $b$ as the end points of the cut and $C_1$ as the contour around them.
We have taken ${\rm Re}(b)>0$ and ${\rm Re}(a)<0$ obeying the assumption (\ref{assumption-weak}).

By regarding the parity symmetry $z \leftrightarrow -z$ of the potential $V(z)$ (\ref{potential-CS}), we assume $a=-b $.
Then, by performing the integrals in (\ref{ansatz-2-cut-resolvent}), we obtain
\begin{align}
v(z)=\frac{n}{ \pi  \lambda } \arctan{\left( \sqrt{\frac{z+b}{z-b}}  \right)} -\frac{n}{4 \lambda } \label{resolvent-two-cut}.
\end{align}
We can confirm that this resolvent satisfies the equation (\ref{eq-resolvent}) with $V'(z)$ given by (\ref{eom-z-weak})\footnote{To see that the resolvent (\ref{resolvent-two-cut}) satisfies the equation (\ref{eq-resolvent}), the relation $\arctan x = \frac{i}{2} \log \left( \frac{i+x}{i-x} \right) $ is useful. }.
Thus the ansatz (\ref{ansatz-2-cut-resolvent}) works even in such a singular potential case too.

Now we take the limit $z \to \infty$ and determine  $b$ so that $v(z)$ satisfies the boundary condition (\ref{boundary-v}). By using the expansion $ \arctan (1+x)  = \frac{\pi}{4} + \frac{ x}{2} +O(x^2) $, we obtain
\begin{align}
v(z)=   \frac{nb}{2 \pi \lambda}\frac{1}{z} + O\left(\frac{1}{z^2}\right), \qquad (z \to \infty).
\end{align}
Hence we fix  $b= 2 \pi \lambda/n $, and we finally obtain
\begin{align}
v(z)=\frac{n}{ \pi  \lambda } \arctan{\left( \sqrt{\frac{n z+2\pi \lambda}{n z-2\pi \lambda}}  \right)} -\frac{n}{4 \lambda }. \label{sol-two}
\end{align}
The result is plotted in figure \ref{fig-one-cut} (left).
From this plot, we can read off the eigenvalue density of the solution.
Interestingly the profile of the density is an ``onion head" shape, and it is quite different from the one-cut solution (\ref{one-cut-rho}) obeying the semi-circle law.
Note that, although the density diverges at $z=0$ as $ \rho(z) \sim \log |z|$,  $\rho(z)$ is correctly normalized $\int  \rho(z) dz=1$.

\subsubsection{Effective potential and free energy}
Through the standard matrix model technique \cite{Marino:2012zq}, we compute the effective potential $V_{\rm  eff}(z)$ which is the potential for the single eigenvalue of the matrix in the two-cut solution (\ref{sol-two}).
For $z > 0$, it becomes
\begin{align}
V_{\rm  eff}(z)
&= \int^z dw \left( V'(w)-2\lambda v(w) \right) \nonumber \\
&=nz-\frac{2nz}{\pi} \arctan{\left( \sqrt{\frac{n z+2\pi \lambda}{n z-2\pi \lambda}}  \right) }- 2  \lambda \log \left( z+\sqrt{z^2-\frac{4 \pi ^2 \lambda^2 }{n^2}} \right) .
\end{align}
The result is plotted in figure \ref{fig-one-cut} (right).
As usual, a plateau appears on the cut.
Thus, although the eigenvalue density has the singularity at $z=0$, the effective potential is not so different from the ordinary solutions in the one matrix models \cite{Marino:2012zq}.

We can also compute the free energy of this solution,
\begin{align}
F^{\rm 2-cut}(\lambda,N)=& \frac{ N^2 n^2 \pi i }{4  \lambda}+\frac{N^2}{\lambda} \left[ \int_{-b}^b dz V(z) \rho(z)-\frac{\lambda}{2} {\rm PV} \int_{-b}^b dz \int_{-b}^b dw \rho(z) \rho(w) \log{(z-w)^2} \right] \nonumber
\\
=&\frac{N^2 n^2  \pi i}{4 \lambda}+ N^2\left[ -\frac{1}{2}- \log{\left( \frac{2\pi \lambda}{n} \right)} \right].
\end{align}
Here the first term is from the original classical potential $u_i^2/4\pi i \lambda $.
Similarly we can calculate the free energy of the one-cut solution (\ref{one-cut-sol}),
\begin{align}
F^{\rm 1-cut}(\lambda,N) = -\frac{N^2}{2} \log{\lambda} +\mathcal{O}(\lambda).
\end{align}
Thus the real part of the free energy of the one-cut solution is lower than the two-cut solution at weak coupling, and it will dominate the path integral (\ref{partition}).

\subsubsection{Other multi-cut solutions}

\begin{figure}
\begin{tabular}{ccc}
\begin{minipage}{0.33\hsize}
\begin{center}
        \includegraphics[scale=0.6]{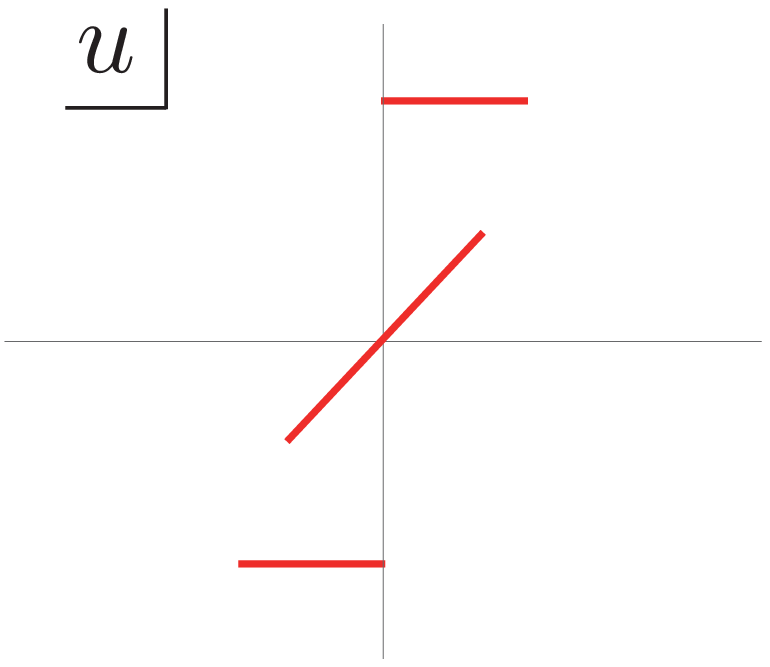}
        multi-cut solution
\end{center}
\end{minipage}
\begin{minipage}{0.33\hsize}
\begin{center}
        \includegraphics[scale=0.6]{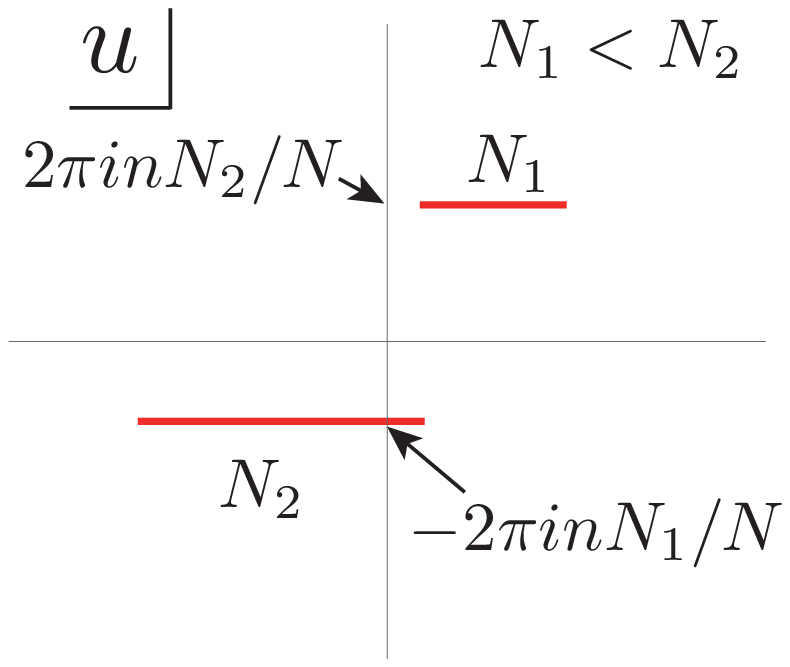}
 asymmetric two-cut solution
\end{center}
\end{minipage}
\begin{minipage}{0.33\hsize}
\begin{center}
        \includegraphics[scale=0.6]{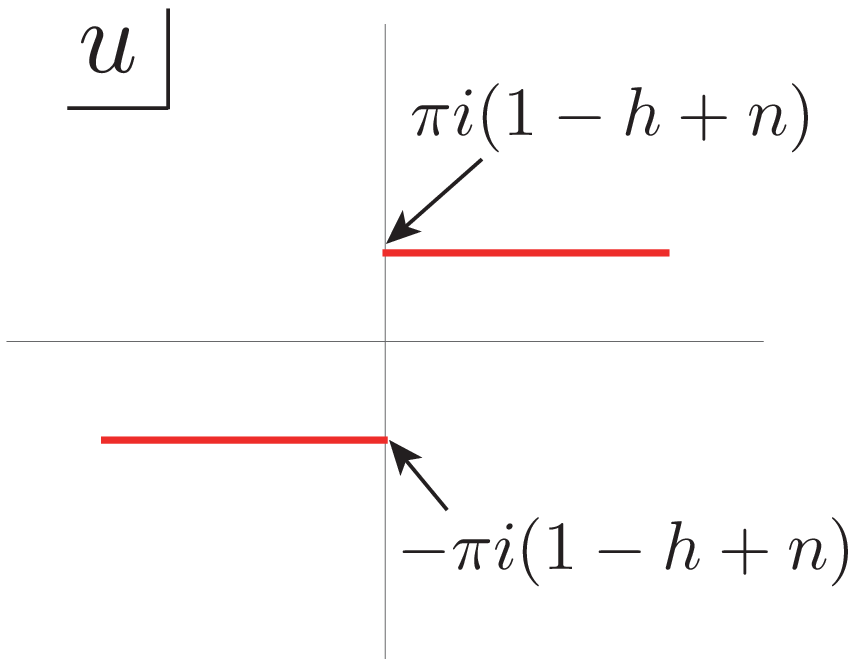}
        adjoint matter
\end{center}
\end{minipage}
\end{tabular}
      \caption{Various multi-cut solutions in the CS matrix models.}
      \label{fig-other-cut}
\end{figure}

We have obtained the two-cut solution inspired by the $N=2$ analysis (\ref{N=2-multi}).
If we change $N$ as $N=3,4,5,\cdots$ in the saddle point equation (\ref{eom-u}), we would easily find various non-trivial solutions.
These solutions indicate the existence of further multi-cut solutions in this system.
For example, at weak coupling, we can approximately superpose the one-cut solution (\ref{one-cut-sol}) and the two-cut solution (\ref{sol-two}), if $n$ is odd. See figure \ref{fig-other-cut} (left).
(When $n$ is odd, the coth interaction between these configurations are negligible.)

Besides, we find the asymmetric two-cut solution in which each cuts consist of $N_1$ and $N_2$$(=N-N_1)$ eigenvalues.
In this solution, the cut for the $N_1$ eigenvalues appears around $u=2\pi i n N_2/N$ and the cut for the $N_2$ eigenvalues does around $u= - 2\pi i n N_1/N$, where $n$ is a non-zero integer. See figure \ref{fig-other-cut} (center).
We show the derivation of the asymmetric two-cut solution at finite $\lambda$ in appendix \ref{app-Exact}.

In this way,  the saddle point equation of the pure CS matrix model allows an infinite number of  multi-cut solutions.
This is a remarkable property of the CS matrix model which is quite distinct from the ordinary one matrix models.
From now we show that similar multi-cut solutions ubiquitously exist in other CS matrix models too.

\section{$\mathcal{N}=2$ Adjoint matter}
\label{sec-Ad}

We consider ${\mathcal N} =2$ supersymmetric CS matter theory coupled to $N_A$ adjoint matters on $S^3$.
Through the localization technique, the partition function of this theory is effectively given by the following matrix model \cite{Jafferis:2010un, Hama:2010av},
\begin{equation}
Z_N^{\rm CSM}(k,h,N_A)=\frac{1}{N!} \int \prod_i \frac{du_i}{2\pi} e^{-\frac{k}{4\pi i} \sum_i u_i^2} \prod_{i<j} \Bigl[ 2 \sinh{\frac{u_i-u_j}{2}} \Bigr]^2 \prod_{i , j} e^{N_A l \left( 1-h+i\frac{u_i-u_j}{2\pi} \right)},
\label{partition-adjoint}
\end{equation}
where $h$ is the $R$-charge of the adjoint matter which takes $0 < h < 1$.
$l(z)$ is the function defined in Ref.\cite{Jafferis:2010un} and it satisfies
\begin{equation}
l'(z)=-\pi z \cot{\pi z}.
\end{equation}
From this partition function, we obtain the saddle point equation 
\begin{align}
u_i
=& \frac{2\pi i \lambda}{N} \sum_{j \neq i} \coth{\frac{u_i-u_j}{2}} \nonumber
\\
& +\frac{ N_A \lambda}{N} \sum_{j \neq i} \left[ \left( \pi(1-h)+i \frac{u_i-u_j}{2} \right) \cot{\left( \pi(1-h)+i \frac{u_i-u_j}{2} \right)} 
\right] \nonumber
\\
& -\frac{ N_A \lambda}{N} \sum_{j \neq i} \left[ \left( \pi(1-h)-i \frac{u_i-u_j}{2} \right) \cot{\left( \pi(1-h)-i \frac{u_i-u_j}{2} \right)}
\right].
\label{eom-adjoint}
\end{align}
As argued in Ref.\cite{Morita:2011cs}, non-trivial multi-cut solutions would exist in this model too.

Here we explore for two-cut solutions.
First we apply the $N=2$ analysis in the pure CS matrix model to the saddle point equation (\ref{eom-adjoint}) too.
Then we find $u_1=-u_2 $ again, and the saddle point equation becomes
\begin{align}
u_1
=& \pi i \lambda  \coth{u_1} +\frac{ N_A \lambda}{2}  \left( \pi( 1-h) +iu_1 \right) \cot{\left( \pi(1-h)+i u_1 \right)} 
 \nonumber
\\
& -\frac{ N_A \lambda}{2}  \left( \pi( 1-h) -iu_1 \right) \cot{\left( \pi(1-h)-i u_1 \right)} .
\end{align}
At weak coupling, we find the solution
\begin{align}
u_1&= \pm \sqrt{\pi i \lambda} + \cdots , \quad
u_1= \pi i n + \frac{\lambda}{n}  + \cdots,  \quad
 u_1= \pm \pi i \left( 1-h +n \right)  \pm \frac{ N_A \lambda}{2} \frac{n}{1-h+n} + \cdots.
\label{N=2-adjoint}
\end{align}
The first and second solutions are the same solutions to the pure CS matrix model at this order, while the last one is new.
Thus a novel class of two-cut solutions associated with this solution may exist in this model.

To find this new two-cut solution at large $N$, we use the ansatz
\begin{align}
u_i
&=   \pi i \left( 1-h +n \right)   +x_i ,  \qquad (i=1, \cdots , N/2),
\nonumber \\
u_{N/2+j}&
= - \pi i \left( 1-h +n \right)  +y_j ,\quad (j=1, \cdots , N/2).
\label{ansatz-2-cut-adjoint}
\end{align}
By substituting them into the saddle point equation (\ref{eom-adjoint}) at weak coupling, we obtain
\begin{align}
 \pi i   \left( 1-h +n \right) 
=& \frac{4\pi i \lambda}{N} \sum_{j \neq i} \frac{1}{x_i-x_j} 
+ \frac{n N_A }{2} \frac{4\pi i \lambda}{N} \sum_{j =1}^{N/2} \frac{1}{x_i-y_j}, \nonumber \\
-  \pi i  \left( 1-h +n \right) 
=& \frac{4\pi i \lambda}{N} \sum_{j \neq i} \frac{1}{y_i-y_j} 
+ \frac{n N_A }{2} \frac{4\pi i \lambda}{N} \sum_{j =1}^{N/2} \frac{1}{y_i-x_j} .
\end{align}
Here we have assumed that $x_i, y_j \sim O(\lambda)$.
Finding the two-cut solution from these equations is still not easy.
However, in the $nN_A=2$ case ($N_A = 2$ and $n = 1$ or $N_A = 1$ and $n = 2$), the equations become almost identical to equation (\ref{eom-weak}), and the solution is given by (\ref{sol-two}) with $n \to 1-h+n$.
 See figure \ref{fig-other-cut} (right) for the schematic plot of this solution.

In additon to this solution, we can perturbatively obtain the two-cut solution associated with 
$u_1= \pi i n +O(\lambda)$  in (\ref{N=2-adjoint}) too.
The solution at the leading order is the same to the pure CS matrix model case (\ref{sol-two}).

Lastly we argue a possible interpretation of the multi-cut solutions in this model.
Recently Ref.\cite{Honda:2016vmv} pointed out that this model may be non-Borel summable in the $1/k$ expansion due to the singularities in the Borel plane.
These singularities are related to the configuration $u_i-u_j= 2\pi i \left( 1-h +n \right)$.
Our two-cut solution (\ref{ansatz-2-cut-adjoint}) indeed satisfy this relation and it indicates that our solutions might be related to the singularities in the Borel plane.
Usually singularities in the Borel plane correspond to some non-perturbative objects, and thus we expect that our solutions might describe such objects in this model.

\section{ Multi-cut solution in ABJM matrix model}
\label{sec-ABJM}

We consider the ABJM matrix model \cite{Kapustin:2009kz}. The partition function of this model is defined by
\begin{align}
&Z_N^{\rm ABJM}(k) \nonumber \\
&= \frac{1}{(N!)^2} \int  \prod_{i=1}^N \frac{d \mu_i}{2\pi}
e^{-\frac{k}{4\pi i}  \mu_i^2 }
 \prod_{j=1}^N \frac{d \nu_i}{2\pi}
e^{\frac{k}{4\pi i}  \nu_j^2 }
   \frac{\prod_{i<j}^{N} \left[ 2\sinh{\frac{\mu_i-\mu_j}{2}} \right]^2 \prod_{i<j}^{N} \left[ 2\sinh{\frac{\nu_i-\nu_j}{2}} \right]^2 }{\prod_{i,j=1}^{N} \left[ 2\cosh{\frac{\mu_i-\nu_j}{2}} \right]^2}.
\end{align}
From this partition function, we obtain the saddle point equation,
\begin{align}
\mu_i =& \frac{2\pi i \lambda}{N} \left[ \sum_{j \neq i}^{N} \coth{\frac{\mu_i-\mu_j}{2}}- \sum_{j=1}^{N} \tanh{\frac{\mu_i-\nu_j}{2}} \right], \nonumber
\\
-\nu_i =& \frac{2 \pi i \lambda}{N} \left[ \sum_{j\neq i}^{N} \coth{\frac{\nu_i-\nu_j}{2}}-\sum_{j=1}^{N} \tanh{\frac{\nu_i-\mu_j}{2}} \right] ,
\label{saddle-ABJM}
\end{align}
where we have defined 't\,Hooft coupling $\lambda\equiv N/k$.
By summing over these saddle point equations, we obtain the relation
\begin{align}
\sum_{i=1}^N \mu_i = \sum_{j=1}^N \nu_j.
\label{ABJM-relation}
\end{align}
In the ABJM matrix model, the parameter $\kappa$ which is related to the 't\,Hooft coupling $\lambda$ by
\begin{align}
\lambda(\kappa) = \frac{\kappa}{8 \pi} {}_3F_2 \left( \frac12, \frac12, \frac12; 1, \frac32; - \frac{\kappa^2}{16} \right)
\label{kappa}
\end{align}
is useful \cite{Drukker:2011zy, Marino:2009jd, Drukker:2010nc}.
Particularly this relation becomes
\begin{align}
\lambda &= \frac{1}{8\pi}  \kappa +O(\kappa^{2}), \qquad  \qquad \qquad \qquad (|\lambda| \ll 1), \\
\lambda &= \frac{1}{2\pi^2} \left(\log \kappa \right)^2 + \frac{1}{24}+O(\kappa^{-2}), \qquad (|\lambda| \gg 1) ,
\label{kappa-lambda}
\end{align}
 at weak and strong coupling, respectively.
 
One solution of the saddle point equation (\ref{saddle-ABJM}) at large $N$ has been derived by 
Drukker, Marino and Putrov \cite{Marino:2009jd, Drukker:2010nc}. 
In this article, we call this solution ``DMP solution".
The spectral curve of the  DMP  solution for $\{\mu_i \}$ is given by 
\begin{align}
Y=e^y, \qquad X=e^x, \qquad Y+ \frac{X^2}{Y}-X^2 + i \kappa X-1=0.
\label{curve-ABJM}
\end{align}
This curve describes two cuts along $[-\alpha,\alpha]$ and $[-\beta+\pi i ,\beta+\pi i]$\footnote{In addition to these two cuts, there are images of them: $[-\alpha+2 \pi in  ,\alpha+2 \pi in]$ and $[ -\beta+\pi i+2 \pi in,\beta+\pi i+2 \pi in]$, ($n \in {\bf Z}$).}.
  (The eigenvalue $\{ \mu_i \}$ and $\{ \nu_i \}$ distribute along the region $[-\alpha,\alpha]$ and $[-\beta,\beta]$, respectively. See figure \ref{fig-ABJM} (left).)
The locations of  $\alpha$ and $\beta$  are determined through the relation
\begin{align}
A= e^\alpha, \qquad B=e^\beta, \qquad A+ \frac{1}{A}+B+\frac{1}{B}=4, 
\qquad 
A+ \frac{1}{A}-B-\frac{1}{B}=2 i \kappa.
\end{align}
By using the relation (\ref{kappa-lambda}), they become
\begin{align}
\alpha &= \sqrt{8 \pi i \lambda}+\cdots, \qquad \beta= i\sqrt{8 \pi i \lambda}+\cdots, \qquad \qquad \qquad \qquad \qquad \qquad \qquad  (|\lambda| \ll 1) , \\
\alpha &= \pi  \sqrt{2 \hat{\lambda} }+ \frac{\pi}{2} i -2i e^{-\pi \sqrt{2 \hat{\lambda}}}+\cdots, \quad \beta= \pi  \sqrt{2  \hat{\lambda}}- \frac{\pi}{2} i+2i e^{-\pi \sqrt{2 \hat{\lambda}}}+\cdots,~~
  (|\lambda| \gg 1),
 \label{ABJM-ab}
\end{align}
where $\hat{\lambda}\equiv \lambda- \frac{1}{24} $. 

\subsection{Multi-cut solution}

\begin{figure}
\begin{center}
        \includegraphics[scale=0.75]{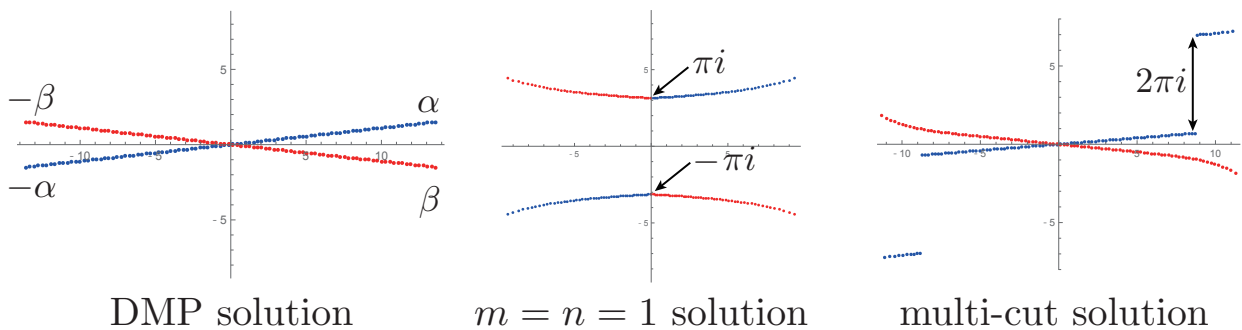}
\end{center}
      \caption{Various solutions in the ABJM matrix model through the numerical analysis.
      We set $N=100$ and $\lambda=10$. 
The blue and red points denote the eigenvalue $\{\mu_i \}$ and $\{\nu_j \}$. 
By changing the initial input of the numerical calculation, we obtain various solutions of the saddle point equation (\ref{saddle-ABJM}).
The left plot corresponds to the DMP solution \cite{ Marino:2009jd, Drukker:2010nc}. 
The center plot corresponds to the ansatz (\ref{ABJM-mn-ansatz}) with $m=n=1$.
The right plot may be related to the D2-brane instantons as we argue in section \ref{sec-ABJM-instanton}.
      }
      \label{fig-ABJM}
\end{figure}

Now we consider finding other solutions.
Similar to the previous CS matrix models, various multi-cut solutions would exist in the ABJM matrix model too.
The simplest solution at weak coupling is the superposition of the solutions of the pure CS matrix model (\ref{sol-two}).
To see this solution, we use the following ansatz
\begin{align}
\mu_i
&=   \pi i n   +x_i   \qquad (i=1, \cdots , N/2), \qquad
\mu_{N/2+j}
= -  \pi i n +y_j \quad (j=1, \cdots , N/2), \nonumber \\
\nu_i &=   \pi i m   +z_i   \qquad (i=1, \cdots , N/2), \qquad
\nu_{N/2+j}
= -  \pi i m +w_j \quad (j=1, \cdots , N/2),
\label{ABJM-mn-ansatz}
\end{align}
where $n$ and $m$ are integers. 
When $n \pm m $ is even, the tanh interaction between $\{x_i, y_j\}$ and $\{w_i, z_j\}$ are negligible  at weak coupling, and the saddle point equation (\ref{saddle-ABJM}) becomes
\begin{align}
\frac{n}{2} & =\frac{2\lambda}{N} \sum_{ j \neq i}^{N/2} \frac1{x_i-x_j} + \frac{2\lambda}{N} \sum_{j =1}^{N/2} \frac1{x_i-y_j},  \qquad (i=1, \cdots , N/2), \nonumber \\
-\frac{n}{2}& =\frac{2\lambda}{N} \sum_{ j \neq i}^{N/2} \frac1{y_i-y_j} + \frac{2\lambda}{N} \sum_{j =1}^{N/2} \frac1{y_i-x_j} ,\qquad (i=1, \cdots , N/2), \nonumber \\
-\frac{m}{2} & =\frac{2\lambda}{N} \sum_{ j \neq i}^{N/2} \frac1{z_i-z_j} + \frac{2\lambda}{N} \sum_{j =1}^{N/2} \frac1{z_i-w_j}  ,\qquad (i=1, \cdots , N/2), \nonumber \\
\frac{m}{2}& =\frac{2\lambda}{N} \sum_{ j \neq i}^{N/2} \frac1{w_i-w_j} + \frac{2\lambda}{N} \sum_{j =1}^{N/2} \frac1{w_i-z_j} ,\qquad (i=1, \cdots , N/2).
\end{align}
These are merely two set of the saddle point equations for the two-cut solutions in the pure CS matrix model (\ref{eom-weak}). ($\lambda \to - \lambda$ for $\{w_i, z_j\}$.)
Thus the solution is given by (\ref{sol-two}).  
See figure \ref{fig-ABJM} for the $m=n=1$ case. (The plot is a numerical result at $N=100$ and $\lambda=10$.)

In addition to this solution, various multi-cut solutions would exist in the ABJM matrix model.
However obtaining analytic solution is not simple generally.
Hence we numerically solve the saddle point equations (\ref{saddle-ABJM}) at finite $N$ and $\lambda$ through the method developed in Refs.\cite{Morita:2011cs, Herzog:2010hf, Niarchos:2011sn, Minwalla:2011ma} in order to find other multi-cut solutions. See figure \ref{fig-ABJM} for some examples.

\subsection{Connection to D2-brane instantons}
\label{sec-ABJM-instanton}

In the DMP solution at strong coupling, two kinds of non-perturbative effects are important: the world sheet instanton and the D2-brane instanton.
Their instanton actions are given by
\begin{align}
\text{world sheet instanton:} \quad 2\pi \sqrt{\lambda}, \qquad
\text{D2-brane instanton:} \quad
 \pi N \sqrt{2/\lambda}.
\label{instanton-factor}
\end{align}
Remarkably, it was pointed out  that the instanton action of the D2-brane instanton is obtained through a sophisticated cycle integral of the spectral curve (\ref{curve-ABJM}) \cite{Drukker:2011zy}.

Here we argue the relation between the multi-cut solution and the D2-brane instanton.
Among the multi-cut solutions we have discussed in the previous section, we focus on the one plotted in figure \ref{fig-ABJM} (right).
This solution can be regarded as a deformation of the DMP solution (figure \ref{fig-ABJM} (left)).
In this solution, the original cut $[-\alpha, \alpha]$ of the DMP solution is divided into three cuts, and the left cut is roughly shifted by $- 2 \pi  i n$ and the right one is shifted by  $ 2 \pi  i n$, where $n$ is a positive integer\footnote{
In our numerical analysis, we could not find the multi-cut solution with negative $n$.
This may be the same mechanism that the cuts in the two-cut solution of the pure CS matrix model (\ref{sol-two}) never appear on the lower right quadrant of the complex plane.
Besides, to retain the relation (\ref{ABJM-relation}), we tune the shifted two cuts symmetric under $\mu \leftrightarrow - \mu$. 
There may be asymmetric solutions also as in the pure CS matrix model case if we tune the cuts appropriately.}.  (Figure \ref{fig-ABJM} is for the $n=1$ case.)
Since the coth interactions between $\{ \mu_j \}$ in the saddle point equation (\ref{saddle-ABJM}) have a periodicity $\mu \to \mu + 2\pi  i $, the interactions between the cuts are still significant even after the $2\pi  i n$ shift, and such a multi-cut configuration can be a solution.

To see the connection to the D2-brane instanton, we investigate the dynamics of the single eigenvalue, say $\mu_N$, in the DMP solution. 
The effective potential for the eigenvalue $\mu_N$ is given by 
\begin{align}
V_{\rm eff} (\mu_N)& =  \frac{N}{4\pi i \lambda}  \mu^2_N + V_{\rm int} (\mu_N) -V_{\rm eff}(\alpha),  
\label{eff-ABJM}
\\
V_{\rm int} (\mu_N) & \equiv
-\sum_{j =1 }^{N-1}
\log \left[ 2\sinh{\frac{\mu_N-\mu_j}{2}} \right]^2+\sum_{j=1}^{N} \log \left[ 2\cosh{\frac{\mu_N-\nu_j}{2}} \right]^2 .
\nonumber
\end{align}
Here we fix $\{ \mu_i \}$ ($i \neq N$) and  $\{ \nu_j \}$  to be the DMP solution and we ignore  the back reaction\footnote{Note that ignoring the back reaction of $\mu_N$ may be subtle, since it breaks the relation (\ref{ABJM-relation}). 
But we can avoid this issue if we set another eigenvalue, say $\mu_1$, as $\mu_1=-\mu_N$. 
This prescription makes the value of the effective action twice.}
 of $\mu_N$.
$-V_{\rm eff}(\alpha)$ in (\ref{eff-ABJM}) is added so that $V_{\rm eff} (\mu_N=\alpha)=0$, since we are interested in the deviation from the DMP solution.

Suppose $\mu_N$ is shifted from  $\mu_N=\alpha$ to $\alpha+ 2\pi  i n$ \footnote{
One question is whether $\mu_N=\alpha+ 2\pi  i n$ is a solution of the saddle point equation for $\mu_N$ derived from the effective potential (\ref{eff-ABJM}) which is given by $0=N \mu_N/2 \pi i \lambda +V'_{\rm int}(\mu_N) $.
Since we have assumed that $\mu_N= \alpha$ is the DMP solution,  it should satisfy  $0=N \alpha/2 \pi i \lambda +V'_{\rm int}(\alpha) $.
However it immediately means that $\mu_N=\alpha+2\pi i n  $ is not a solution due to the periodicity $V'_{\rm int}(\mu +2\pi  i n )=V'_{\rm int}(\mu  )$.
(The same result can be seen through the equation $y=0$, where $y$ is the curve (\ref{curve-ABJM}).)
On the other hand, the multi-cut solutions in the numerical analysis indicates the existence of the solution near $\mu_N=\alpha+2\pi i n $.
It implies that the back reaction to other eigenvalues are important to construct the solution.
In this paper, we do not consider the back reaction and just evaluate the value of the effective potential (\ref{eff-ABJM}) at  $\mu_N=\alpha+2\pi i n $, since the correction from the back reaction to the value of the effective potential may be higher order in the $1/N$ expansion. 
On the other hand, the back reaction may correct the total free energy.
It would be an interesting future work to find the multi-cut solution analytically by taking into account the back reaction and evaluate the free energy.
}.
Then this configuration can be regarded as a single point version of the multi-cut solution discussed above.
In this case, the effective potential becomes\begin{align}
V_{\rm eff} (\alpha+ 2\pi  i n)& =  \frac{N}{4\pi i \lambda}  (\alpha+ 2\pi i n )^2 + V_{\rm int} (\alpha+ 2\pi  i n) -V_{\rm eff}(\alpha) \nonumber \\
&= n \frac{  N  \alpha}{\lambda}  - i  \frac{N \pi n^2 }{\lambda} \nonumber \\
&= n  N  \pi \sqrt{2 /\lambda}+ \cdots, \qquad (|\lambda| \gg 1). 
\end{align}
Here we have used the periodicity of the interaction $ V_{\rm int} (\alpha+ 2\pi  i n)=V_{\rm int} (\alpha)$, and equation (\ref{ABJM-ab}).
Therefore the value of the effective potential with $n=1 $ precisely agrees with the instanton action of the D2-brane instanton at strong coupling $|\lambda| \gg 1$ (\ref{instanton-factor}).
This quantitative agreement indicates that the multi-cut solution may be interpreted as a condensation of the D2-brane instantons.

\section{Discussions}
\label{sec-discussion}

We have revealed that the CS matrix models in the 't\,Hooft limit have remarkable properties and an infinite number of multi-cut solutions exit.
These solutions might illuminate non-perturbative aspects of the CS theories and string theory.
However, as mentioned in the introduction,  we just evaluated the saddle point equations of the matrix models, and we should understand whether these solutions contribute to the path integral or not.
On the other hand, we have seen some possible connections between the multi-cut solutions and other known non-perturbative objects: the singularities in the Borel plane in the CS matter theory \cite{Honda:2016vmv} and the D2-brane instantons in the ABJM theory \cite{Drukker:2011zy}.
Hence some of the multi-cut solutions might be indeed physical.
To develop understanding the role of the multi-cut solutions, it would be important to compare our results and other non-perturbative properties of the CS matrix models further.

In the pure CS matrix model (\ref{partition}), the following instanton factor has been obtained through an A-cycle integral \cite{Pasquetti:2009jg},
\begin{align}
\frac{N}{2\pi i \lambda}\left(
4\pi^2 \lambda n - 4 \pi^2 n m \right),
\label{Instanton-Pas}
\end{align}
where $n$ and $m$ are integers which are concerned with the periodicity of coth in the saddle point equation (\ref{eom-u}).
Thus the periodicity of coth is crucial in both this instanton and our multi-cut solutions.
However this instanton is obtained through the A-cycle integral, and the instanton factor (\ref{Instanton-Pas}) may not be related to the free energy of the multi-cut solution.

In the ABJM matrix model, so-called Fermi gas approach is quite powerful \cite{Marino:2011eh}. 
By using this technique, various important results on the membrane instantons have been obtained \cite{Hatsuda:2012dt, Calvo:2012du, Hatsuda:2013gj, Hatsuda:2013oxa}.
(See a review \cite{Hatsuda:2015gca}.)
Thus comparing these results with our multi-cut solutions may be important.
However, in order to do this quantitatively, we need to develop our computation technique and obtain the analytic expressions for the multi-cut solutions.

Another important direction of investigation is finding the gravity duals of the multi-cut solutions through the AdS/CFT correspondence.\\

Lastly we argue possible instanton factors in the CS matrix models.
In section \ref{sec-ABJM-instanton}, we have seen that the instanton factor of the D2-brane instanton at strong coupling is reproduced by evaluating the difference of the effective potential through the shift of the eigenvalue $\alpha \to \alpha + 2\pi i $, 
where $\alpha$ is the end point of the cut.
Therefore it might be possible to extract  similar instanton factors in other CS matrix models through the $2 \pi i $ shift of the eigenvalue at the end point of the cut. 
Particularly, if the interaction terms have the periodicity $u_i \to u_i + 2\pi i $ as in the pure CS and ABJM case, the instanton factor arises from the classical action $V_{\rm classical}(u)$ only, and it becomes
\begin{align}
V_{\rm classical}(\alpha+ 2 \pi  i )-V_{\rm classical}(\alpha),
\end{align}
where $\alpha$ is the end point of the cut in a suitable solution of the saddle point equation\footnote{This instanton factor differs from (\ref{Instanton-Pas}) in the pure CS matrix model, since this factor is related to B-cycle whereas (\ref{Instanton-Pas}) is related to A-cycle.}.
Hence once we know the location of the end point of the cut, we may easily obtain the instanton factor\footnote{In the adjoint matter case (\ref{partition-adjoint}),  the term $ \sum_{i,j} l(1-h+i (u_i-u_j)/2 \pi )$ induces the additional cuts at $2\pi i (\pm h+n) $ ($n \in {\mathbf R}$).
Thus there might be instantons corresponding to the shift of the eigenvalue $\alpha \to \alpha + 2\pi i (\pm h+n) $.}.
It may be valuable to explore these instantons in the CS matrix models.

\paragraph{Acknowledgements}
The authors would like to thank Yasuyuki Hatsuda, Masazumi Honda, Kazumi Okuyama, Takao Suyama and Asato Tsuchiya for valuable discussions and comments.
The work of T.~M. is supported in part by Grant-in-Aid for Scientific Research (No. 15K17643) from JSPS.

\appendix

\section{Exact two-cut solution at finite $\lambda$ in pure CS matrix model}
\label{app-Exact}

We calculate the  two-cut solutions in the pure CS matrix model at large $N$ with a finite $\lambda$.
In section \ref{sec-CS}, we assumed that each cut is composed by $N/2$ eigenvalues.
Here we relax this condition and consider the two cuts composed by $N_1$ and $N_2=N-N_1$ eigenvalues.
(See figure \ref{fig-other-cut} (center).)

By summing the saddle point equation (\ref{eom-u}), we obtain a relation
\begin{align}
\sum_{i=1}^N u_i=0.
\label{constraint-CS}
\end{align}
By regarding this equation, we use the following ansatz for the two-cut solution
\begin{align}
u_i
&=  2 \pi i n  \frac{N_2}{N} +x_i ,  \qquad (i=1, \cdots , N_1),
\nonumber \\
u_{N_1+j}&
= - 2 \pi i n  \frac{N_1}{N}+y_j ,\quad (j=1, \cdots , N_2),
\label{ansatz-asym}
\end{align}
where $n$ is an integer.
We can choose this integer $n$ positive without loss of generality.
By substituting this ansatz into the saddle point equation (\ref{eom-u}), we obtain
\begin{align}
  \frac{N_2}{N} n+\frac{1}{2\pi i} x_i & =\frac{\lambda}{N} \sum_{ j \neq i}^{N_1} \coth{\frac{x_i-x_j}{2}} + \frac{\lambda}{N} \sum_{j =1}^{N_2} \coth{\frac{x_i-y_j}{2}}, \qquad (i=1, \cdots , N_1), \nonumber \\
-  \frac{N_1}{N}n+\frac{1}{2\pi i} y_i & =\frac{\lambda}{N} \sum_{ j \neq i}^{N_2} \coth{\frac{y_i-y_j}{2}} + \frac{\lambda}{N} \sum_{j =1}^{N_1} \coth{\frac{y_i-x_j}{2}} ,\qquad (i=1, \cdots , N_2).
\label{saddle-CS}
\end{align}
In order to solve these equations, we use the following assumption
\begin{align}
{\rm Re}\left( y_j \right)  \le {\rm Re}\left( z_0 \right) \le {\rm Re}\left( x_i \right),
 \label{assumption-weak-2}
\end{align}
where $z_0$ is a complex number representing the boundary of $\{ x_i \}$ and $\{ y_j \}$.
This assumption is the same to the assumption (\ref{assumption-weak}) if $z_0=0$.
Since the system is asymmetric if $N_1 \neq N_2$, the boundary $z_0$ may be non-zero in general.
Thanks to this assumption, we can combine the saddle point equations (\ref{saddle-CS}) into the single equation by using the variable $z_i = x_i$ for $1 \le i \le N_1$ and $z_{i+N_1} = y_i$ for $1 \le i \le N_2$,
\begin{align}
2n \theta \left({\rm Re}\left( z -z_0 \right)\right) - \frac{N_1}{N} 2n + \frac{1}{\pi i} z_i  & =  \frac{2 \lambda}{N} \sum_{ j \neq i}^{N} \coth{\frac{z_i-z_j}{2}}.
\label{eom-z}
\end{align}
If the first two terms on the left hand side did not exist, the problem is solving the saddle point equation in the original pure CS matrix model (\ref{eom-u}).
Hence we can apply the technique developed there \cite{Aganagic:2002wv, Halmagyi:2003ze, Suyama:2016nap}.
We define new variable $Z_i \equiv e^{z_i}$ and $Z_0 \equiv e^{z_0}$, and rewrite equation (\ref{eom-z}) as
 \begin{align}
U'(Z_i) & =  \frac{2 \lambda}{N} \sum_{ j \neq i}^{N} \frac{Z_i+Z_j}{Z_i-Z_j}  , \qquad
U'(Z)
\equiv 2n \theta \left( |Z| - |Z_0|  \right) - \frac{N_1}{N} 2n + \frac{1}{\pi i} \log Z .
\label{eom-Z}
\end{align}
Now we introduce the eigenvalue density $\rho(Z)$ and resolvent $v(Z)$ as
\begin{align}
\rho(Z) \equiv \frac{1}{N} \sum_i^N \delta(Z-Z_i) , \qquad v(Z)=\int_A^B dZ' \rho(Z') \frac{Z+Z'}{Z-Z'},
\label{def-resolvent}
\end{align}
where $A$ and $B$ are the locations of the end points of the eigenvalue distribution, which satisfy $0<|A| \le |Z_0|$ and $|Z_0| \le |B|$ because of the assumption (\ref{assumption-weak-2}).
By using this resolvent, we rewrite the saddle point equation (\ref{eom-Z}) as
\begin{align}
U'(Z)= \lambda \lim_{\epsilon \to 0} \left( v(Z+ i \epsilon)+v(Z- i \epsilon) \right).
\label{eom-v}
\end{align}
Therefore the equations which we want to solve are similar to those of the Hermitian matrix model. 
The difference is only the boundary condition of the resolvent.
From the definition of the resolvent (\ref{def-resolvent}), it should satisfy
\begin{align}
v(Z) \to 1 \qquad (Z \to \infty), \qquad v(Z) \to -1 \qquad (Z \to 0).
\label{boundary-v-Z}
\end{align}
Note that the eigenvalue density is now derived as 
\begin{align}
\rho(Z)
=& -\frac{1}{4\pi i Z} \lim_{\epsilon \rightarrow 0} \left[ v(Z+i\epsilon)-v(Z-i\epsilon) \right] ,\qquad (Z \in [A,B] ~) ,
\label{density}
\end{align}
through (\ref{def-resolvent}).

As in the analysis at the weak coupling, we apply the ``one-cut" ansatz to the resolvent,
\begin{align}
v(Z)=&\frac{1}{\lambda} \int_{C_1} \frac{dZ'}{4\pi i} \frac{U'(Z')}{Z-Z'} \sqrt{\frac{(Z-A)(Z-B)}{(Z'-A)(Z'-B)}} \nonumber \\
=& \frac{1}{2\pi i \lambda} \int_{Z_0}^B dZ' \frac{2n}{Z-Z'} \sqrt{\frac{(Z-A)(Z-B)}{(Z'-A)(Z'-B)}} 
-\frac{1}{4\pi i \lambda} \int_{C_1} dZ' \frac{\frac{N_1}{N} 2n}{Z-Z'} \sqrt{\frac{(Z-A)(Z-B)}{(Z'-A)(Z'-B)}} \nonumber
\\
&+\frac{1}{4\pi i \lambda} \frac{1}{\pi i} \int_{C_1} dZ' \frac{ \log{Z'}}{Z-Z'} \sqrt{\frac{(Z-A)(Z-B)}{(Z'-A)(Z'-B)}},
\label{ansatz-v}
\end{align}
where the integral contour $C_1$  goes around the cut$:[A,B]$ in a counter-clockwise way.
Then we can calculate each integrals as
\begin{align}
\int_{Z_0}^B dZ' \frac{1}{Z-Z'} \sqrt{\frac{(Z-A)(Z-B)}{(Z'-A)(Z'-B)}}
=& 2i \arctan{\left( \sqrt{\frac{Z-A}{Z-B}} \sqrt{\frac{B-Z_0}{Z_0-A}} \right)},
\nonumber \\
\int_{C_1} dZ' \frac{1}{Z-Z'} \sqrt{\frac{(Z-A)(Z-B)}{(Z'-A)(Z'-B)}}
=& 2\pi i , \nonumber \\
\int_{C_1} dZ' \frac{ \log{Z'}}{Z-Z'} \sqrt{\frac{(Z-A)(Z-B)}{(Z'-A)(Z'-B)}}
=&4\pi i \log{\left( \frac{Z+\sqrt{AB}-\sqrt{(Z-A)(Z-B)}}{\sqrt{A}+\sqrt{B}} \right)}.
\end{align}
Hence the resolvent becomes
\begin{align}
v(Z)
=& \frac{2n}{\pi  \lambda } \arctan{\left( \sqrt{\frac{Z-A}{Z-B}} \sqrt{\frac{B-Z_0}{Z_0-A}} \right)} -\frac{n}{ \lambda } \frac{N_1}{N} \nonumber \\
& 
+\frac{1}{\pi i \lambda} \log{\left( \frac{Z+\sqrt{AB}-\sqrt{(Z-A)(Z-B)}}{\sqrt{A}+\sqrt{B}} \right)}.
\label{resolvent-sol}
\end{align}
We can confirm that this resolvent satisfies the equation (\ref{eom-v}) as in the weak coupling case. 

\begin{figure}
\begin{center}
        \includegraphics[scale=0.7]{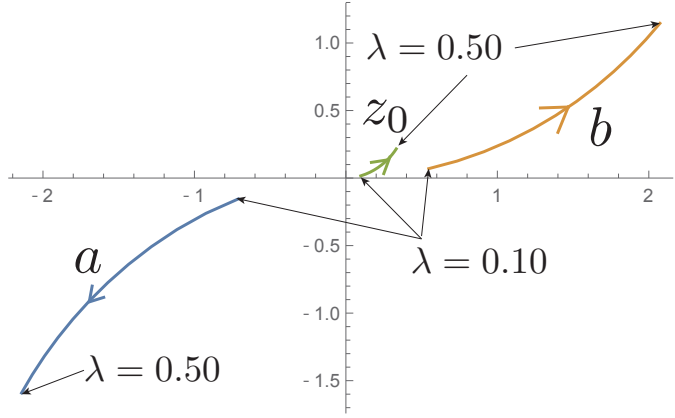}
\end{center}
        \caption{Numerical result for $a=\log A$, $b=\log B$ and $z_0=\log Z_0$ with $N_1/N_2=2/3$ and $n=1$.
        We change $\lambda$ from $0.10$ to $0.50$ and plot $a$, $b$ and $z_0$ by solving equation (\ref{bd-cond}) and (\ref{C2-cond}).
Note that the cuts in the $u$-plane appear along $[-4\pi i/5 +a,-4\pi i/5 +z_0]$ and $[6\pi i/5 +z_0,6\pi i/5 +b]$ via (\ref{ansatz-asym}).     }
        \label{fig-N1-N2}
\end{figure}

Now we determine the constant $A$, $B$ and $Z_0$.
Firstly, the solution (\ref{resolvent-sol}) has to satisfy the boundary condition (\ref{boundary-v-Z}) which is evaluated as
\begin{align}
& 2n  \arctan\left( \sqrt{\frac{B-Z_0}{Z_0-A}}\right) - \pi n \frac{N_1}{N} -i \log\left( \frac{\sqrt{A}+\sqrt{B}}{2}\right)=\pi \lambda , \nonumber \\
&2n \arctan\left(\sqrt{\frac{A}{B}} \sqrt{\frac{B-Z_0}{Z_0-A}}\right)  - \pi n \frac{N_1}{N} 
+ i\log\left( \frac{\sqrt{A}+\sqrt{B}}{2 \sqrt{AB}}\right)=-\pi \lambda .
\label{bd-cond}
\end{align}
In addition, the numbers of the eigenvalues on the cut $[Z_0,B]$ and  $[A,Z_0]$ should be $N_1$ and $N_2$ respectively.
Through the expression of $\rho(Z)$ (\ref{density}), this condition becomes
\begin{align}
\frac{N_1}{N} = \int_{Z_0}^B dZ~ \rho(Z) = \frac{1}{4\pi i} \int_{C_2} dZ ~\frac{v(Z)}{Z},
\label{C2-cond}
\end{align}
where the integral contour $C_2$  goes around the cut$:[Z_0,B]$ in a counter-clockwise way\footnote{Through (\ref{density}), $\rho(Z)$ on the cut$:[Z_0,B]$ becomes
\begin{align}
\rho(Z)= \frac{1}{2 \pi^2  \lambda Z} \left( n \log\left[ \frac{ 1 \pm \sqrt{\frac{Z-A}{B-Z}} \sqrt{\frac{B-Z_0}{Z_0-A}} }{1 \mp \sqrt{\frac{Z-A}{B-Z}} \sqrt{\frac{B-Z_0}{Z_0-A}} } \right]
-\log{\left( \frac{Z+\sqrt{AB}+i \sqrt{(Z-A)(B-Z)}}{\sqrt{A}+\sqrt{B}} \right)}
+ \frac{1}{2} \log Z
 \right).
\end{align}
Here we have used the relation
\begin{align}
\left( \frac{Z+\sqrt{AB}+i \sqrt{(Z-A)(B-Z)}}{\sqrt{A}+\sqrt{B}} \right)
\left(  \frac{Z+\sqrt{AB}-i \sqrt{(Z-A)(B-Z)}}{\sqrt{A}+\sqrt{B}}\right) =Z .
 \end{align}
 }.
Note that the condition for the cut$:[A,Z_0]$  is automatically satisfied if these conditions are satisfied.
From these conditions, we can fix the constant $A$, $B$ and $Z_0$ at least numerically.
We plot the result for the $N_1:N_2=2:3$ and $n=1$ case in figure \ref{fig-N1-N2}.

\subsection{Weak coupling analysis}

At weak coupling, we can solve equation (\ref{bd-cond}) and (\ref{C2-cond}) perturbatively in $\lambda$.
At the leading order, the boundary condition (\ref{bd-cond}) can be solved
\begin{align}
 b =z_0+\frac{2\pi \lambda}{n} \tan{\left( \frac{\pi}{2} \frac{N_1}{N} \right)}
+O(\lambda^2),
 \nonumber
\\
a =z_0-\frac{2\pi \lambda}{n} \tan{\left( \frac{\pi}{2} \frac{N_2}{N} \right)}+O(\lambda^2),
\end{align}
where $a \equiv \log A$ and $b \equiv \log B$.
Now we need to fix $z_0$ by solving (\ref{C2-cond}).
Through equation (\ref{constraint-CS}) and (\ref{ansatz-asym}), equation (\ref{C2-cond}) becomes the following relation
\begin{align}
\int_a^b dz \,z \rho(z)=0.
\end{align}
 We can perturbatively solve this equation, and obtain
\begin{align}
z_0=
-\frac{\lambda \pi }{2n} 
\left( \tan{\left( \frac{\pi}{2} \frac{N_1}{N} \right)}-\tan{\left( \frac{\pi}{2} \frac{N_2}{N} \right)} \right) +O(\lambda^2).
\end{align}
Therefore, if, for example, $N_1<N_2$, $z_0$ becomes positive as shown in figure \ref{fig-other-cut} (center).
This is consistent with the numerical result shown in figure \ref{fig-N1-N2}.

\subsection{$N_1=N_2$ at finite $\lambda$}
In the case of $N_1=N_2$, due to the symmetry $z \leftrightarrow -z$, $Z_0=1$ and $A=1/B$ would be satisfied.
(They imply $z_0=0$ and $a=-b$ in the original $z$ variable.)
Then the condition (\ref{bd-cond}) and (\ref{C2-cond}) reduce to 
\begin{align}
\left( \frac{i-\sqrt{B}}{i+\sqrt{B}}
\right)^n \frac{\sqrt{B}+\frac{1}{\sqrt{B}}}{2} = \exp\left(\pi i \left(\lambda +\frac{n}{2}\right)\right).
\end{align}
Here we have  used the relation $\arctan x = \frac{i}{2} \log \left( \frac{i+x}{i-x} \right) $.
We can analytically solve this equation for $n=1,2,3$ and $4$. 
For example, the solution for the $n=1$ case is given by
\begin{align}
\sqrt{B} = -i \left( e^{\pi i \lambda}-1
\right) + \sqrt{2  e^{\pi i \lambda}-  e^{2\pi i \lambda}}.
\end{align}
Here we have chosen the solution which satisfies $|B|>|Z_0|=1$.

 \bibliographystyle{JHEP} 
 \bibliography{CS} 

\end{document}